\documentclass[aps,ams,reprint, superscriptaddress]{revtex4-2}
\usepackage{amsmath,  amsfonts}
\usepackage{algorithmic}
\usepackage{graphicx}
\usepackage{appendix}
\usepackage{amssymb}
\usepackage{hyperref}
\usepackage{physics}
\usepackage{booktabs}
\usepackage{todonotes}
\usepackage{siunitx}
\usepackage{xcolor}
\begin{document}
\title{Symmetry Adapted Analysis of Screw Dislocation: \\ Electronic Structure and Carrier Recombination Mechanisms in GaN}
\author{Yuncheng Xie}
\thanks{These authors are co-first authors and contributed equally to this work.}
\affiliation{Department of Physics, Fudan University, Shanghai 200433, China}
\affiliation{Key Laboratory of Computational Physical Sciences (Ministry of Education), State Key Laboratory of Surface Physics,
Fudan University, Shanghai 200433, China}

\author{Haozhe Shi}
\thanks{These authors are co-first authors and contributed equally to this work.}
\affiliation{Department of Physics, Fudan University, Shanghai 200433, China}
\affiliation{Key Laboratory of Computational Physical Sciences (Ministry of Education), State Key Laboratory of Surface Physics,
Fudan University, Shanghai 200433, China}

\author{Menglin Huang}
\affiliation{Key Laboratory of Computational Physical Sciences (Ministry of Education), State Key Laboratory of Surface Physics,
Fudan University, Shanghai 200433, China}
\affiliation{College of Integrated Circuits and Micro-Nano Electronics, Fudan University, Shanghai 200433, China}

\author{Weibin Chu}
\affiliation{Department of Physics, Fudan University, Shanghai 200433, China}
\affiliation{Key Laboratory of Computational Physical Sciences (Ministry of Education), State Key Laboratory of Surface Physics,
Fudan University, Shanghai 200433, China}

\author{Shiyou Chen}
\affiliation{Key Laboratory of Computational Physical Sciences (Ministry of Education), State Key Laboratory of Surface Physics,
Fudan University, Shanghai 200433, China}
\affiliation{College of Integrated Circuits and Micro-Nano Electronics, Fudan University, Shanghai 200433, China}

\author{Xin-Gao Gong}
\affiliation{Department of Physics, Fudan University, Shanghai 200433, China}
\affiliation{Key Laboratory of Computational Physical Sciences (Ministry of Education), State Key Laboratory of Surface Physics,
Fudan University, Shanghai 200433, China}
\date{\today}

\begin{abstract}
    As fundamental one-dimensional defects, screw dislocations profoundly reshape the energy landscape and carrier dynamics of crystalline materials. By restoring the exact algebra of the screw dislocation group, we unveil the latent symmetry constraints that govern the electronic structure, providing a more rigorous physical picture than the conventional treatments. When applied to GaN, the method yields a band-connectivity constraint and rigorous dipole selection rules
    for polarization-resolved transitions. Combined with computed Hamiltonian
    matrix, the approach gives symmetry-filtered radiative and dielectric calculations and reveals a
    piezoelectrical effect at the dislocation core that strongly suppresses radiative recombination. The pronounced dominance of non-radiative capture over radiative recombination highlights the detrimental impact of screw dislocations on the luminous efficiency of GaN, providing a theoretical foundation for optimizing dislocation-limited optoelectronic devices.
\end{abstract}

\maketitle

\section{Introduction}
Screw dislocations are prevalent in semiconductor materials and can significantly affect material properties, particularly electronic processes such as radiative and nonradiative carrier recombination~\cite{shockley1952statistics}. Understanding these mechanisms is important for the design of next-generation semiconductor devices~\cite{zhao2023trap, zhang2022origin, stoneham1981non, liang2023origin, valero2024estimating, wang2023low}. 

Since a screw dislocation breaks the translational symmetry perpendicular to the dislocation axis, the conventional Bloch theorem no longer directly applies. Therefore previous studies have typically employed large supercells and performed density functional theory (DFT) calculation~\cite{gao2025first, wang2020large, jia2019parallel, ye2021large}. While the supercell method remains the standard for modeling dislocation geometries, traditional DFT treatments often view these large ensembles as mere disordered clusters, thereby failing to exploit the underlying screw symmetry inherent in the structure. Consequently, the rich physical information remains locked within an unnecessarily large and opaque Hamiltonian. This paper presents a framework that, while still utilizing the supercell geometry, explicitly recovers and leverages this hidden symmetry to block-diagonalize the electronic structure and unveil novel mathematical features~\cite{sharma2022symmetry, sharma2021real}.

In this paper, we first present a rigorous theoretical analysis of the symmetry properties inherent to screw dislocations. This framework establishes a symmetry-adapted basis that allows the Hamiltonian to be block-diagonalized, with each block corresponding to a specific irreducible representation. Consequently, the total band structure is resolved into independent components labeled by these representations.  Leveraging this decomposition, we reveal hidden interconnections between bands originating from different symmetry channels. Beyond the electronic structure, we derive the dipole selection rules for polarization-resolved optical transitions. These rules facilitate the calculation of the radiative recombination coefficient, which is subsequently compared with its non-radiative counterpart. Our results provide critical insights and theoretical guidance for future experimental investigations into the optoelectronic properties of dislocations.

\section{Symmetry Adapted Basis}
In this section we develop a concise, self-contained theoretical framework for exploiting the n-fold screw symmetry
that characterizes the systems studied in this work. We begin by constructing symmetry-adapted basis functions using the plane-wave approach. Due to the limitations of this method for screw-symmetric systems, we then introduce a localized atomic orbital framework. This construction yields basis functions that, as eigenstates of the screw operator, can effectively give a block-diagonalized Hamiltonian. 

For generality we consider a n-fold screw operation
that rotates by $2\pi/n$ about the $z$-axis and translates by $m c/n$ along $z$. So a screw symmetry is characterized by the rotation index $n$ and translation index $m$.

\subsection{Screw operator and its representations}
Let $S$ denote the n-fold screw operator acting on spatial coordinates by
\begin{equation}
S:\ (x,y,z) \mapsto (x\cos\theta-y\sin\theta, x\sin\theta+y\cos\theta,z+mc/n),
\end{equation}
where $\theta=2\pi/n$. The Hamiltonian $H$ of the crystal
commutes with $S$ in the idealized symmetry-preserving model, i.e. $[H,S]=0$. For Bloch states labeled by the
crystal momentum $k_z$ along the $z$-axis the eigenvalues $\lambda$ of $S$ satisfy
\begin{equation}
\lambda^n=e^{ik_z mc}.
\end{equation}
It is convenient to parametrize the n branches by an integer index $\mu\in\{0,1,\dots,n-1\}$:
\begin{equation}\label{eq:lambda_mu}
\lambda_{\mu}(k_z)=e^{i(k_z\frac{mc}{n}+\mu\frac{2\pi}{n})}.
\end{equation}

\subsection{Plane-wave expansion}
Consider the standard plane-wave ansatz used in supercell electronic-structure calculations,
\begin{equation}
\psi_{k_z,\mathbf{G}_\parallel,\ell}(\mathbf{r})=e^{ik_z z}e^{i\mathbf{G}_\parallel\cdot\mathbf{r}_\parallel},
\end{equation}
where $\mathbf{r}_\parallel=(x,y)$ and $\mathbf{G}_\parallel$ is an in-plane reciprocal-lattice vector (set by
the supercell). To construct eigenstates of symmetry operator one must form linear combinations of plane waves
whose in-plane wavevectors are related by the $C_n$ rotation. For each orbit of n rotated wavevectors
$\{\mathbf{G}_\parallel,R_z(\theta)\mathbf{G}_\parallel,\dots\}$ one can form symmetry-adapted linear combinations
that transform according to the representations in \eqref{eq:lambda_mu},
\begin{equation}
    \phi_{\mathbf{G}_{\|}, l, \mu}(\mathbf{r}) = e^{i k_z z} \sum_{j=0}^{n-1} \left[ e^{i (\mu-l) \frac{2\pi}{n}} \right]^j e^{i \mathbf{G}_j \cdot (x, y)} e^{i \mathbf{G}_l z}
\end{equation}
where $\mathbf{G}_j$ are the rotated $\mathbf{G}_{\|}$ vectors for $j$ times and $\mathbf{G}_l$ are the reciprocal lattice vectors along the screw axis ($z$-direction) associated with the index $l$ .

While the plane-wave method is effective for standard electronic structure calculations, it poses significant challenges for the explicit symmetry analysis pursued in this work. Specifically, the symmetry-based decomposition requires the construction and manipulation of the Hamiltonian matrix. Given the large number of plane waves (determined by a fine in-plane $\mathbf{G}_\parallel$ grid), the resulting Hamiltonian matrix becomes exceedingly large. This high dimensionality makes the subsequent block-diagonalization and analytical interpretation computationally prohibitive.

These considerations motivate the construction of a more compact basis set that remains adapted to the action of $S$. Localized atomic orbitals, when combined with Bloch sums along $z$ and appropriate screw-symmetrization, provide an ideal framework for this purpose, offering a much more manageable matrix dimension for symmetry analysis.

\subsection{Localized atomic orbitals}
We assume a set of localized orbitals $\{\lvert R,\alpha\rangle\}$ centered at atomic sites $R=(x_R,y_R,z_R)$, where
$\alpha$ labels orbital type (e.g. site index within the projected unit cell and orbital angular momentum). The
atomic positions are invariant under translation by $c$ along $z$ and under the screw $S$ (in the sense that for each
site $R$ the site $SR$ is also present). Because translation by $c$ is a symmetry we may form Bloch sums along $z$:
\begin{equation}\label{eq:blochz}
\lvert R,\alpha;k_z\rangle \equiv \frac{1}{\sqrt{N_z}}\sum_{n\in\mathbb{Z}}e^{ik_z z_{R,n}}\,\lvert R+n c\hat{z},\alpha\rangle,
\end{equation}
where $z_{R,n}=z_R+n c$ and $N_z$ is the number of cells along $z$ used for normalization. These states carry the
Bloch phase $e^{ik_z c}$ under $T_z(c)$ and remain localized in the projected $xy$-plane.

Sites in a given $C_n$ orbit can be labeled by a representative index $p$ and an internal index $j\in\{0,1,\dots,n-1\}$ such
that $R_{p,j}=S^j R_{p,0}$. Denote the corresponding Bloch-summed orbitals by $\lvert p,j;\alpha;k_z\rangle$.

We now form screw-symmetry-adapted linear combinations with localized atomic orbital symmetrization. For each orbit $p$, orbital label
$\alpha$, Bloch momentum $k_z$ and representation index $\mu$ define
\begin{equation}\label{eq:chi_def}
\lvert\chi_{p,\alpha,k_z,\mu}\rangle \equiv \frac{1}{\sqrt{n}}\sum_{j=0}^{n-1}e^{-ij\mu\frac{2\pi}{n}}\,U_j^{(\alpha)}\,\lvert p,j;\alpha;k_z\rangle,
\end{equation}
where $U_j^{(\alpha)}$ denotes the action of the planar rotation $R_z(j2\pi/n)$ on the orbital degrees of freedom
(for scalar $s$ orbitals $U_j^{(\alpha)}=1$). The coefficients $e^{-ij\mu\frac{2\pi}{n}}$ implement projection onto the
one-dimensional representation labeled by $\mu$.

\subsection{Proof: $\lvert\chi_{p,\alpha,k_z,\mu}\rangle$ is the eigenstate of $S$}
To validate \eqref{eq:chi_def} is actually the eigenstate of screw operator $S$, we act with $S$ on \eqref{eq:chi_def}. Using the geometric action $S\,R_{p,j}=R_{p,j+1}$ and the fact that
z-coordinates obey $z_{p,j+1}=z_{p,j}+\frac{mc}{n}$, together with the rotational action $S\,U_j^{(\alpha)}=U_{j+1}^{(\alpha)}$,
one finds
\begin{align}
S\lvert\chi_{p,\alpha,k_z,\mu}\rangle &= \frac{1}{\sqrt{n}}\sum_{j=0}^{n-1} e^{-ij\mu\frac{2\pi}{n}} S\,U_j^{(\alpha)}\lvert p,j;\alpha;k_z\rangle \nonumber\\
&= \frac{1}{\sqrt{n}}\sum_{j=0}^{n-1} e^{-ij\mu\frac{2\pi}{n}} U_{j+1}^{(\alpha)}\,e^{ i k_z \frac{mc}{n} }\lvert p,j+1;\alpha;k_z\rangle \nonumber\\
&= e^{ik_z \frac{mc}{n}}\,e^{i\mu\frac{2\pi}{n}}\,\lvert\chi_{p,\alpha,k_z,\mu}\rangle.
\end{align}
Therefore $\lvert\chi_{p,\alpha,k_z,\mu}\rangle$ is an eigenstate of $S$ with eigenvalue $\lambda_{\mu}(k_z)$ given in
\eqref{eq:lambda_mu}. The key ingredients are the explicit Bloch factors in \eqref{eq:blochz} that use the true
atomic $z$-coordinates and the $C_n$ projection phases $e^{-ij\mu\frac{2\pi}{n}}$.

\subsection{Block-diagonalization of the Hamiltonian}
Since $\lvert\chi_{p,\alpha,k_z,\mu}\rangle$ is an eigenstate of the screw operator $S$, the commutativity $[H, S] = 0$ dictates that the Hamiltonian $H$ can be block-diagonalized in accordance with the principles of group theory. Consider the matrix element between two symmetry-adapted
states with representation indices $\mu$ and $\mu'$,
\begin{equation}
H_{\mu\mu'}\equiv\langle\chi_{p,\alpha,k_z,\mu}\rvert H \lvert\chi_{q,\beta,k_z,\mu'}\rangle.
\end{equation}
Because $S$ commutes with $H$ we can insert $S^{-1}S$ adjacent to $H$ and use that $S\lvert\chi_{\cdot,\mu}\rangle=
\lambda_{\mu}\lvert\chi_{\cdot,\mu}\rangle$ to obtain
\begin{equation}
H_{\mu\mu'} = \lambda_{\mu}^*\lambda_{\mu'}\,H_{\mu\mu'}.
\end{equation}
Hence
\begin{equation}
(1-\lambda_{\mu}^*\lambda_{\mu'})H_{\mu\mu'}=0.
\end{equation}
If $\lambda_{\mu}\neq\lambda_{\mu'}$ (which holds for $\mu\neq \mu'$ for generic $k_z$) we conclude $H_{\mu\mu'}=0$. Thus matrix
elements between different screw representations vanish and the Hamiltonian is block-diagonal in the basis
$\{\lvert\chi_{p,\alpha,k_z,\mu}\rangle\}$. 

In practice, we can construct the projection operators to block-diagonalize the original Hamiltonian. Define the $k$-dependent projector onto the $\mu$-th screw representation by the standard discrete Fourier projector
in the group algebra:
\begin{equation}\label{eq:projector}
P_{\mu}(k)=\frac{1}{n}\sum_{j=0}^{n-1}\lambda_{\mu}(k)^{-j}\,\hat{S}^j,
\end{equation}
where the eigenvalues $\lambda_\mu$ are defined as~\eqref{eq:lambda_mu} and for any state $\lvert\psi\rangle$ with $\hat{S}\lvert\psi\rangle=\lambda_{\mu}(k)\lvert\psi\rangle$ the projector
acts as $P_{\mu}(k)\lvert\psi\rangle=\lvert\psi\rangle$. Therefore the Hamiltonian decomposes as
\begin{equation}\label{eq:block_decomp}
\hat{H} = \sum_{\mu=0}^{n-1} P_{\mu}(k)\,\hat{H}\,P_{\mu}(k) = \bigoplus_{\mu=0}^{n-1} H_\mu(k).
\end{equation}
Each block $H_\mu(k)=P_{\mu}(k)\hat{H}P_{\mu}(k)$ acts only within the $\mu$-th symmetry subspace and may be diagonalized independently, yielding
bands $\epsilon_{n,\mu}(k_z)$ that are classified by the screw representation.

\section{Electronic structure}

\subsection{Crystal structure and computational method}
We employ the single 6-atom-ring full-core model~\cite{matsubara2013properties, matsubara2014threading}, a standard configuration in GaN screw dislocation calculations, as a test case for our study. The dislocation core is located at the center of a hexagonal GaN ring, characterized by a Burgers  vector $\mathbf{b}=[0001]c$ , where $c$ denotes the structural repetition parameter along the $z$-direction. This configuration exhibits a $6_2$ screw symmetry with respect to the dislocation center, which means the rotation index $n=6$ and translation index $m=2$.
Therefore, the generator of the screw symmetry is taken to be
\begin{equation}
\hat{S}=\{C_6\mid c/3\},
\end{equation}
where $C_6$ is a rotation by $2\pi/6$ about the axis and the translation component is $c/3$ along the axis.

\begin{figure}[ht]
\centering
\includegraphics[width=1.0\columnwidth]{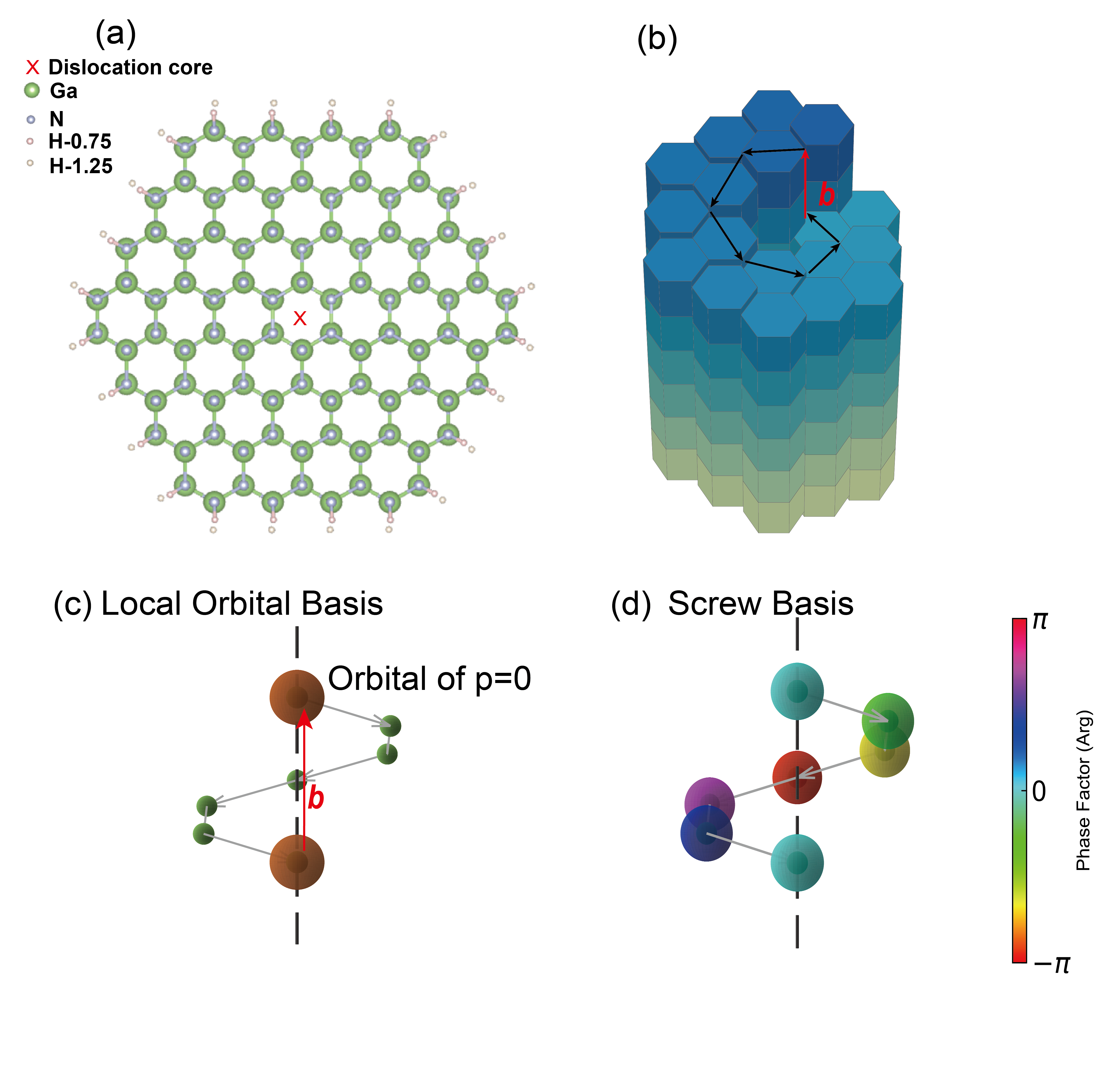}
\caption{Structural models of a screw-dislocation GaN nanowire. (a) Cross-sectional structure of the nanowire passivated by fractional H atoms. (b) Schematic illustration of the screw dislocation symmetry. (c, d) Schematic illustration of constructing screw-symmetry-adapted basis functions from localized atomic orbitals. The visualization depicts an $s$-orbital model at $k_z=0$ for the screw representation index $\mu=1$. }
\label{fig:stru}
\end{figure}

Given that the geometry of a screw dislocation inherently breaks translational symmetry, we adopt a quasi-one-dimensional GaN nanowire model. Vacuum layers are applied in the $xy$-plane to isolate the wire, and the lateral surfaces are passivated using pseudo-hydrogen atoms with fractional charges to eliminate dangling bond surface 
states.\cite{gao2025first} 

% To facilitate the symmetry-adapted analysis, the supercell is constructed to enclose complete symmetry orbits, ensuring that for any included atom, all its images generated by the screw operator are also present. 

Specifically, the initial wurtzite lattice is constructed using experimental parameters ($a=3.189$\AA, $c=5.185$\AA\cite{vurgaftman2003band}, $u=0.375$), and the dislocation is introduced via the continuum displacement field $u_{z}(x,y)=\frac{|\mathbf{b}|}{2\pi}\arctan(y/x)$. For surface passivation, pseudo-hydrogen atoms ($Z=1.25e$ for H$_{\text{Ga}}$, $Z=0.75e$ for H$_{\text{N}}$) are placed to satisfy tetrahedral coordination. The position of each passivator is defined by the missing bond direction $\hat{\mathbf{n}}_i \propto - \sum_{j} (\mathbf{r}_j - \mathbf{r}_i)$, calculated as the normalized negative vector sum of the surface atom's existing bonds. This deterministic placement method rigorously preserves the $6_2$ screw axis symmetry, as it is strictly defined by the intrinsic local bonding geometry. Consequently, the constructed supercell successfully encloses complete symmetry orbits, ensuring that for any included atom, all its images generated by the screw operator are also present.
The resulting atomic structure is illustrated in Fig.~\ref{fig:stru}.

DFT calculations are performed using the ABACUS package to obtain the Hamiltonian in a localized atomic orbital representation~\cite{chen2010systematically, li2016large, lin2024ab}. This Hamiltonian is subsequently transformed and diagonalized within the symmetry-adapted subspaces. To ensure the accuracy of localized state bandgaps, we employed the HSE06 hybrid functional for calculations~\cite{lin2020efficient, lin2020accuracy}.

\subsection{Hamiltonian and Band structure}
\begin{figure}[ht]
\centering
\includegraphics[width=1.0\columnwidth]{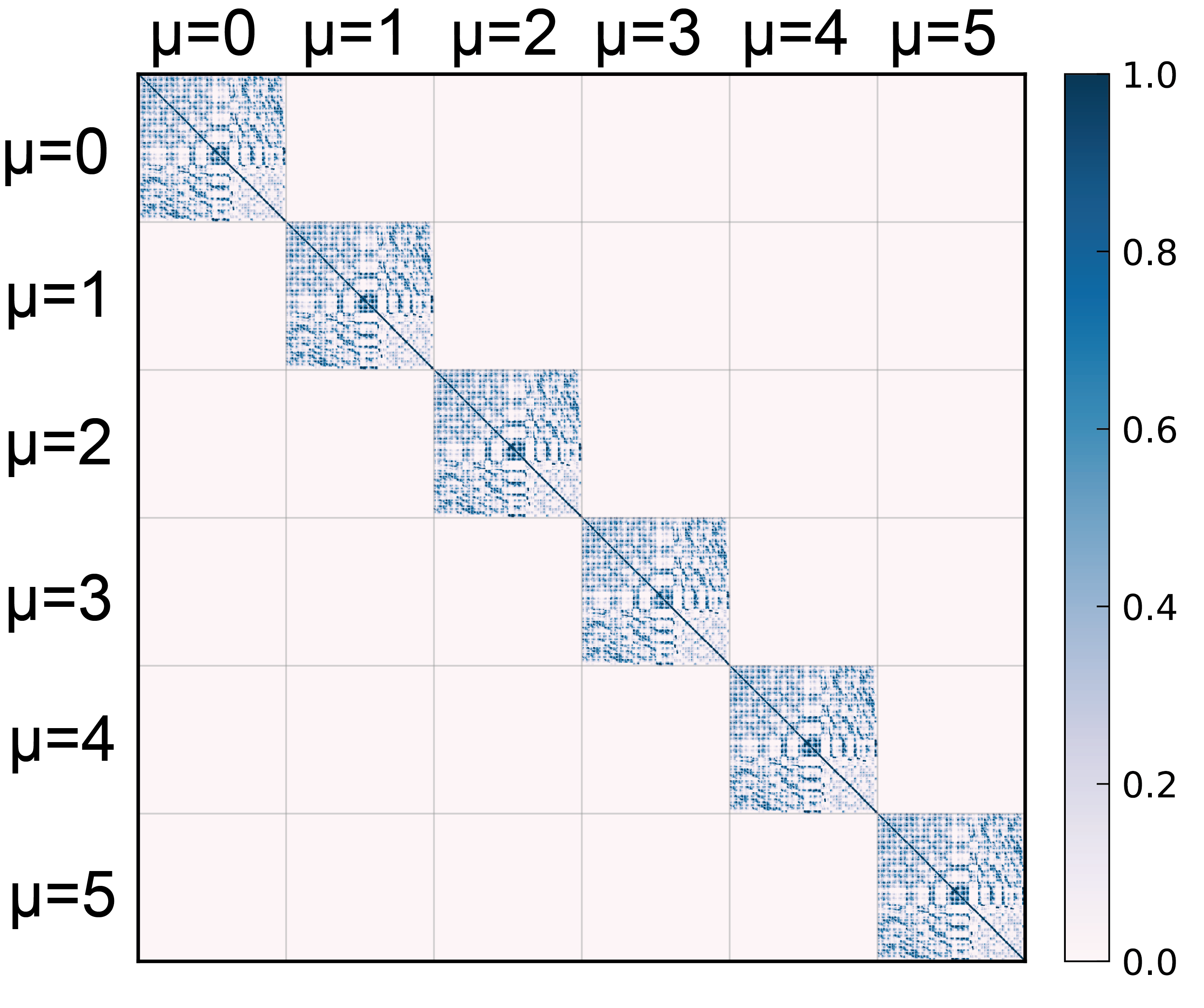}
\caption{Heatmap of the Hamiltonian matrix in the localized-orbital basis. The color represents the magnitude of the matrix elements. After transformation to the screw-symmetric basis the matrix decomposes into six independent blocks $H_\mu(k)$. This result presents the meaning of the symmetry-adapted basis.}
\label{fig:hamiltonian_heatmap}
\end{figure}

Theoretical analysis based on Eq.~\eqref{eq:block_decomp} indicates that the Hamiltonian for GaN ($n=6$) decomposes into six distinct blocks. This block-diagonal structure is clearly visualized in the heatmap of the Hamiltonian within the localized-orbital basis (Fig.~\ref{fig:hamiltonian_heatmap}) once the projectors $P_{\mu}(k)$ are applied. Independently diagonalizing each block yields the symmetry-resolved dispersion relations $\{\epsilon_{n,\mu}(k)\}_{n}$ for $\mu = 0, \dots, 5$. The composite band structure, representing the union of these six sub-spectra, can be rendered as a color-coded plot that highlights the dominant $\mu$-character of each branch.

\begin{figure}[ht]
\centering
\includegraphics[width=1.05\columnwidth]{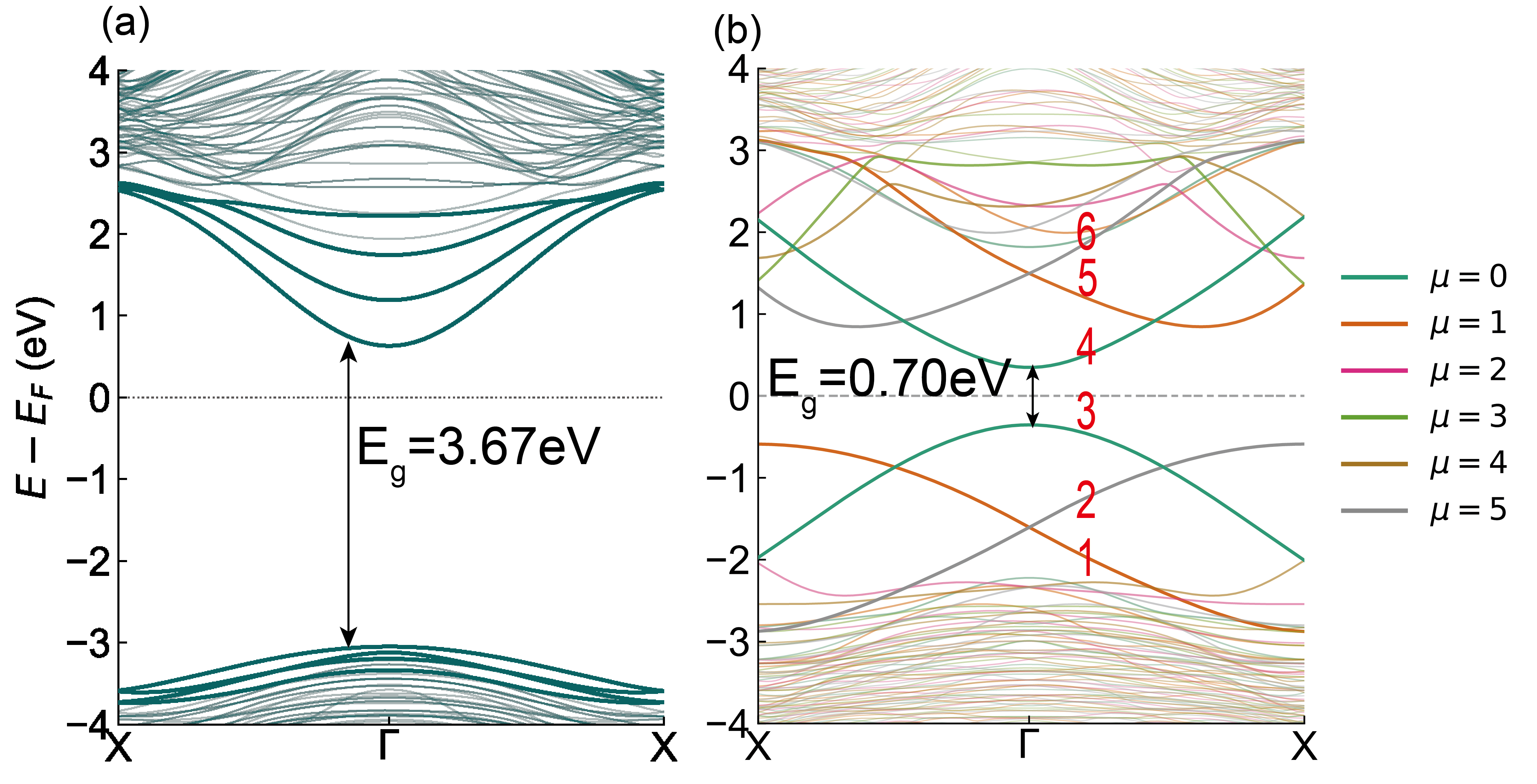}
\caption{Calculated band structures of GaN nanowires. (a) The ideal nanowire. (b) The screw-dislocated nanowire, where the dislocation-induced gap states are labeled from $1$ to $6$. The bands are color-coded according to their screw symmetry index $\mu$.}
\label{fig:band}
\end{figure}

Fig.~\ref{fig:band} compares the band structures of ideal and screw-dislocated GaN nanowires. The pristine nanowire exhibits a band gap of 3.68 eV, in good agreement with experimental values (3.51 eV \cite{monemar1974fundamental, vurgaftman2003band}). Conversely, the introduction of a screw dislocation induces six distinct localized states and significantly reduces the band gap to 0.70 eV, which is consistent with prior calculations \cite{gao2025first} and aligns with the $\sim$0.75 eV deep-level defects experimentally observed in GaN via optical deep-level transient spectroscopy (ODLTS) \cite{sui2024emission}.

The emergence of these discrete states is primarily driven by the localized potential well at the screw dislocation core, which "traps" electronic states within the bandgap. This strong localization effectively narrows the electronic gap to 0.70 eV. A more detailed physical analysis of how the resulting spatial charge separation further impacts recombination processes is provided in Section~\ref{physical_interpretation}.

Utilizing our symmetry-based analysis, we demonstrate that these localized states originate from the $\mu=0, 1,$ and $5$ symmetry channels. Notably, the states flanking the Fermi level (labeled 3 and 4) are exclusively associated with the $\mu=0$ representation.

\subsection{Band connectivity and coupling chains}
We now prove the band-flow relation that constrains how bands from different $\mu$-blocks connect across the
Brillouin-zone boundary. Let the axial reciprocal-lattice vector be $G=2\pi/c$. Under the translation $k\to
k+G$ the phase in $\lambda_\mu$ changes as
\begin{equation}
    \begin{split}
        \lambda_{\mu}(k+G) &= \exp\!\Bigl[i\Bigl(\frac{(k+G)c}{3}+\frac{\pi\mu}{3}\Bigr)\Bigr] \\
        &= \exp\!\Bigl[i\Bigl(\frac{kc}{3}+\frac{\pi\mu}{3}+\frac{2\pi}{3}\Bigr)\Bigr]
        \\
        &=\exp\!\Bigl[i\Bigl(\frac{kc}{3}+\frac{\pi(\mu+2)}{3}\Bigr)\Bigr]=\lambda_{\mu+2}(k),
    \end{split}
\end{equation}
where the index $\mu+2$ is understood modulo six. Therefore the screw-eigenvalue of a state at $(k+G)$ equals the
screw-eigenvalue of a state in block $\mu+2$ at $k$. Energy bands are continuous functions of $k$, hence an
eigenvalue branch originating in block $\mu$ at wave vector $k$ must connect to an eigenvalue branch in block
$\mu+2$ upon translation by $G$. This establishes the band-flow rule
\begin{equation}\label{eq:delta_mu}
\Delta\mu=+2\quad(\mathrm{mod}\;6)\quad\text{for }k\mapsto k+G.
\end{equation}

\begin{figure}[ht]
\centering
\includegraphics[width=1.0\columnwidth]{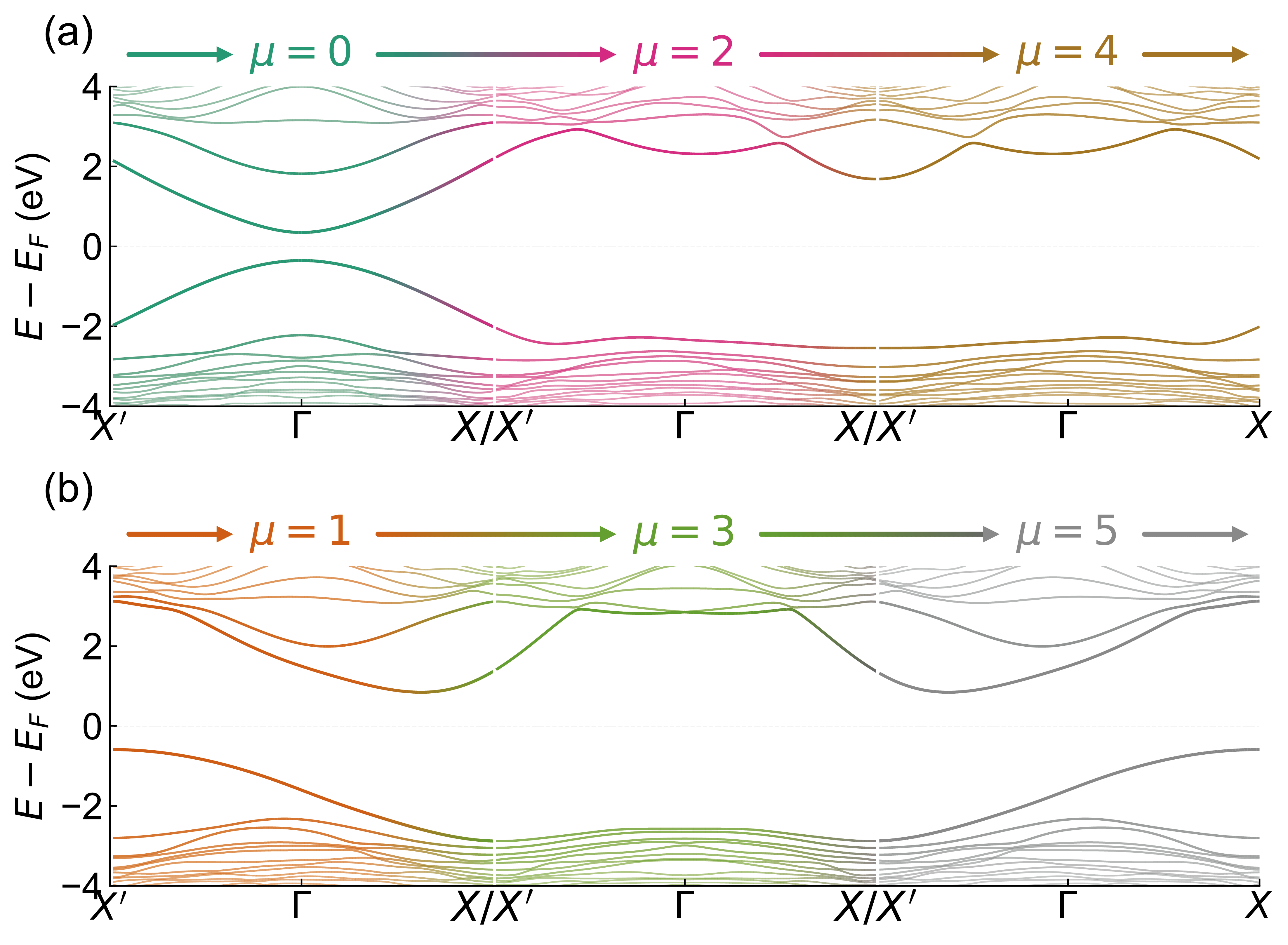}
\caption{Six blocks split into two independent band-flow chains: (a)the even chain
$(\,0\to2\to4\to0\,)$; (b)the odd chain $(\,1\to3\to5\to1\,)$. This rule is the internal connection between bands from different blocks under screw symmetry. }
\label{fig:band_structure_mu}
\end{figure}

Consequently, as the Fig.~\ref{fig:band_structure_mu} shows, the six blocks split into two independent band-flow chains: the odd chain
$(\,1\to3\to5\to1\,)$ and the even chain $(\,0\to2\to4\to0\,)$ (indices taken modulo six with an appropriate
choice of origin). These coupling chains are mathematical constraints imposed by the screw symmetry and the
associated phase accumulation; they are independent of microscopic Hamiltonian details and hence robust.

\section{Radiative recombination}
Having established the screw-symmetric decomposition of Hamiltonian and the
symmetry-resolved bands $\epsilon_{n,\mu}(k)$ in the preceding section, we now derive the optical selection
rules and compute radiative observables for GaN systems with sixfold screw symmetry.

\subsection{Dipole operator and screw transformation}
In the electric-dipole approximation the light-matter interaction is governed by the dipole operator
\(\mathbf{d}=-e\mathbf{r}\). Under the screw operator $S=\{C_6\mid c/3\}$ the spatial operators transform as
\begin{equation}\label{eq:S_on_rp}
S\,\mathbf{r}\,S^{-1}=R_z(2\pi/6)\,\mathbf{r}+\mathbf{\tau},\qquad S\,\mathbf{p}\,S^{-1}=R_z(2\pi/6)\,\mathbf{p},
\end{equation}
with $\mathbf{\tau}=(0,0,c/3)$. For transition matrix elements between distinct eigenstates the constant translation
term does not contribute, so $\mathbf{d}$ is treated as a rank-1 tensor under rotations.

It is convenient to decompose the dipole into spherical components
\(d^{(m)}\) with $m\in\{0,\pm1\}$:
\begin{equation}\label{eq:dipole_components}
d^{(0)}=-e z,\qquad d^{(\pm1)}= -\frac{e}{\sqrt{2}}(x\pm i y),
\end{equation}
which transform under $C_6$ as
\begin{equation}
S\,d^{(m)}\,S^{-1}=e^{i 2\pi m/6}\,d^{(m)}.
\end{equation}

\subsection{Selection rules}
Let $|\psi_{i,k}^{(\mu_i)}\rangle$ and $|\psi_{f,k}^{(\mu_f)}\rangle$ be eigenstates of $\hat{H}$ and $S$ at the same
axial wave vector $k$, with screw eigenvalues $\lambda_{\mu_i}(k)$ and $\lambda_{\mu_f}(k)$ defined in
\eqref{eq:lambda_mu}. Consider the dipole matrix element
\begin{equation}
M_{fi}^{(m)}(k)=\langle\psi_{f,k}^{(\mu_f)}\,|\,d^{(m)}\,|\,\psi_{i,k}^{(\mu_i)}\rangle.
\end{equation}
Insert the identity $S^{-1}S$ and use the transformation properties to obtain
\begin{align}
M_{fi}^{(m)}(k) &= \langle\psi_{f,k}^{(\mu_f)}|S^{-1}S\,d^{(m)}\,S^{-1}S|\psi_{i,k}^{(\mu_i)}\rangle \nonumber\\
&= \lambda_{\mu_f}(k)^*\,\lambda_{\mu_i}(k)\,e^{i2\pi m/6}\,M_{fi}^{(m)}(k).
\end{align}
Nonzero matrix elements therefore require
\begin{equation}\label{eq:selection_raw}
\lambda_{\mu_f}(k)^*\,\lambda_{\mu_i}(k)\,e^{i2\pi m/6}=1.
\end{equation}
Using \eqref{eq:lambda_mu} the $k$-dependent phase cancels (photon momentum is neglected in the dipole
approximation), yielding the compact selection rule
\begin{equation}\label{eq:selection_rule}
\Delta\mu\equiv\mu_f-\mu_i \equiv m \pmod{6}.
\end{equation}
Thus axial polarization ($m=0$) connects states within the same block, while transverse circular polarizations
connect blocks differing by $\pm1$ (mod 6). This finding aligns with and further refines the general group-theoretic predictions, with the derived selection rules summarized in Table~\ref{tab:selection_rules}.
\begin{table}[htbp]
\centering
\caption{Optical selection rules in screw-dislocated GaN.} 
\label{tab:selection_rules}
\begin{tabular}{lcll}
\toprule
\textbf{Polarization} & \textbf{$m$} & \textbf{Selection Rule} & \textbf{Interpretation} \\
\midrule
Axial ($\bm{E} \parallel \hat{z}$) & 0 & $\mu_f = \mu_i$ & Intra-block only \\
Circular $\sigma^+$ & $+1$ & $\mu_f \equiv \mu_i + 1$ & Screw momentum $+1$ \\
Circular $\sigma^-$ & $-1$ & $\mu_f \equiv \mu_i - 1$ & Screw momentum $-1$ \\
Linear $\bm{E} \perp \hat{z}$ & $\pm 1$ & $\mu_f \equiv \mu_i \pm 1$ & Mixed $\pm 1$ channels \\
\bottomrule
\end{tabular}
\end{table}

\begin{figure}[ht]
\centering
\includegraphics[width=1.0\columnwidth]{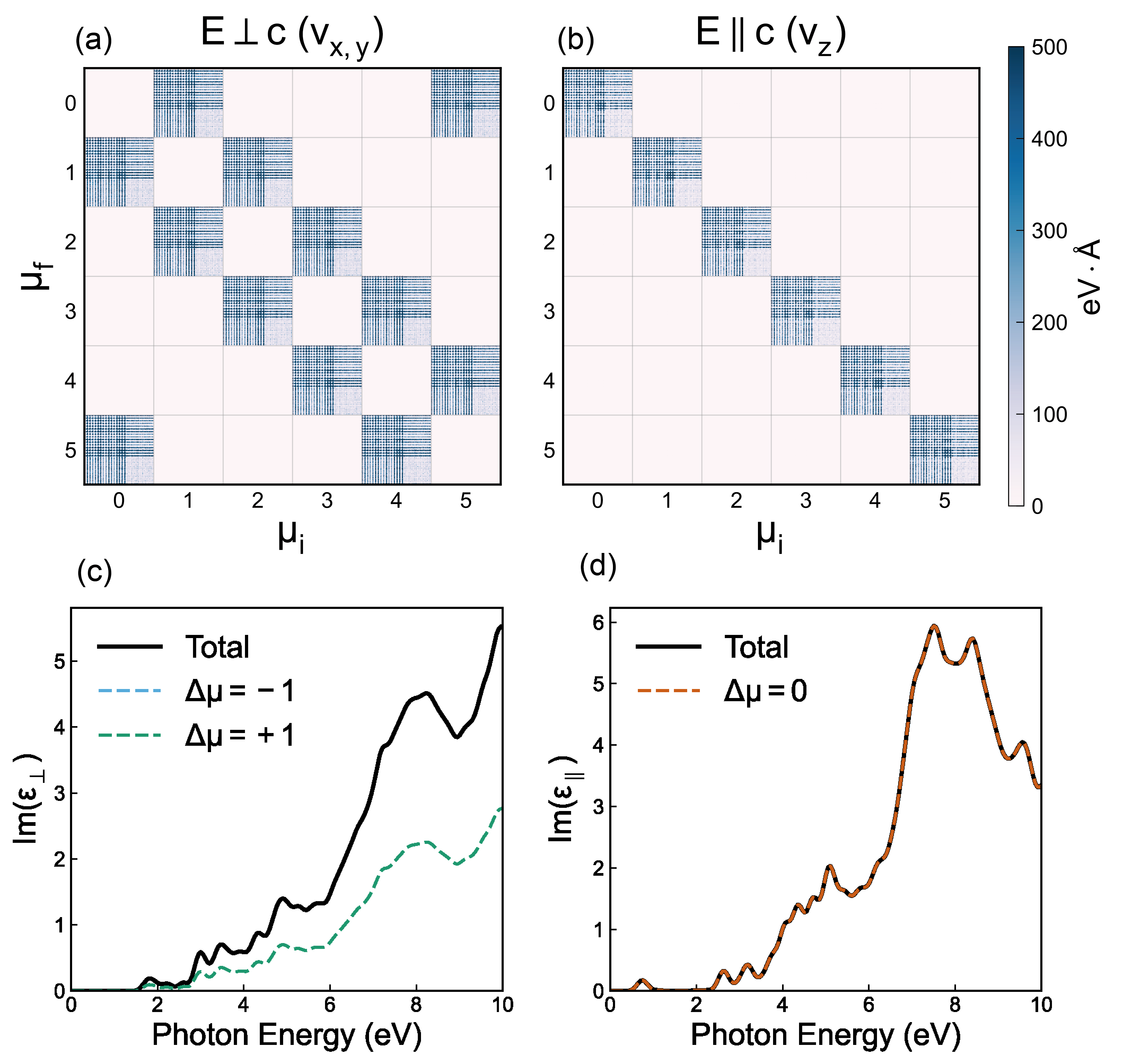}
\caption{Optical properties of screw-dislocated GaN nanowires.
(a, b) Optical transition matrix elements in the screw basis for transverse (perpendicular to $c$) and axial (parallel to $c$) polarizations, respectively. (c, d) Imaginary parts of the dielectric function for the corresponding directions, showing the decomposition into different $\Delta \mu$ channels.}
\label{fig:selection_rules}
\end{figure}

The calculated optical matrix elements in the screw basis are shown in Figs.~\ref{fig:selection_rules}(a) and (b), with the corresponding $\Delta \mu$-resolved imaginary parts of the dielectric function provided in (c) and (d). The results confirm our selection rules: transverse components ($v_x, v_y$) mediate transitions restricted to $\Delta\mu = \pm 1$, whereas the axial component ($v_z$) is confined to the $\Delta\mu = 0$ channel due to its block-diagonal nature.

\subsection{Radiative Recombination Rate}

We calculate the band-to-band radiative recombination rate $R_{\mathrm{rad}}$ based on the theory of spontaneous emission within the electric dipole approximation.\cite{wang2022effective} For a system with refractive index $n_r$, the total spontaneous emission rate per unit volume at an injected carrier density $n$ is given by:

\begin{equation}\label{eq:R_rad}
\begin{split}
    R_{\mathrm{rad}} &= \frac{n_r e^2}{3\pi\varepsilon_0 m_0^2 c^3 \hbar^2 V} \\
    &\times \sum_{k} w_k \sum_{c \in \mathrm{CB}} \sum_{v \in \mathrm{VB}} f_c(1-f_v)\, (E_{ck} - E_{vk})\, \lvert \mathbf{p}_{cv}(k) \rvert^2,
\end{split}
\end{equation}
where $V$ is the effective volume determined by the nanowire cross-section,  $w_k$ are the $k$-point weights, and $E_{ck}$ ($E_{vk}$) are the eigenenergies of the conduction (valence) band states. The Fermi-Dirac occupation factors $f_{c}$ and $f_{v}$ are determined by the electron and hole quasi-Fermi levels, $E_{Fn}$ and $E_{Fp}$, which are solved self-consistently to satisfy the charge neutrality condition $n = \frac{1}{V}\sum_{k,c} f_c = \frac{1}{V}\sum_{k,v} (1-f_v)$.

Utilizing the screw Bloch basis, we decompose $R_{\mathrm{rad}}$ into symmetry channels defined by $\Delta \mu = (\mu_c - \mu_v) \pmod 6$. The radiative coefficient is then given by $B_{\mathrm{rad}} = R_{\mathrm{rad}} / n^2$, which characterizes the optical polarization anisotropy.

\begin{figure}[ht]
\centering
\includegraphics[width=1.0\columnwidth]{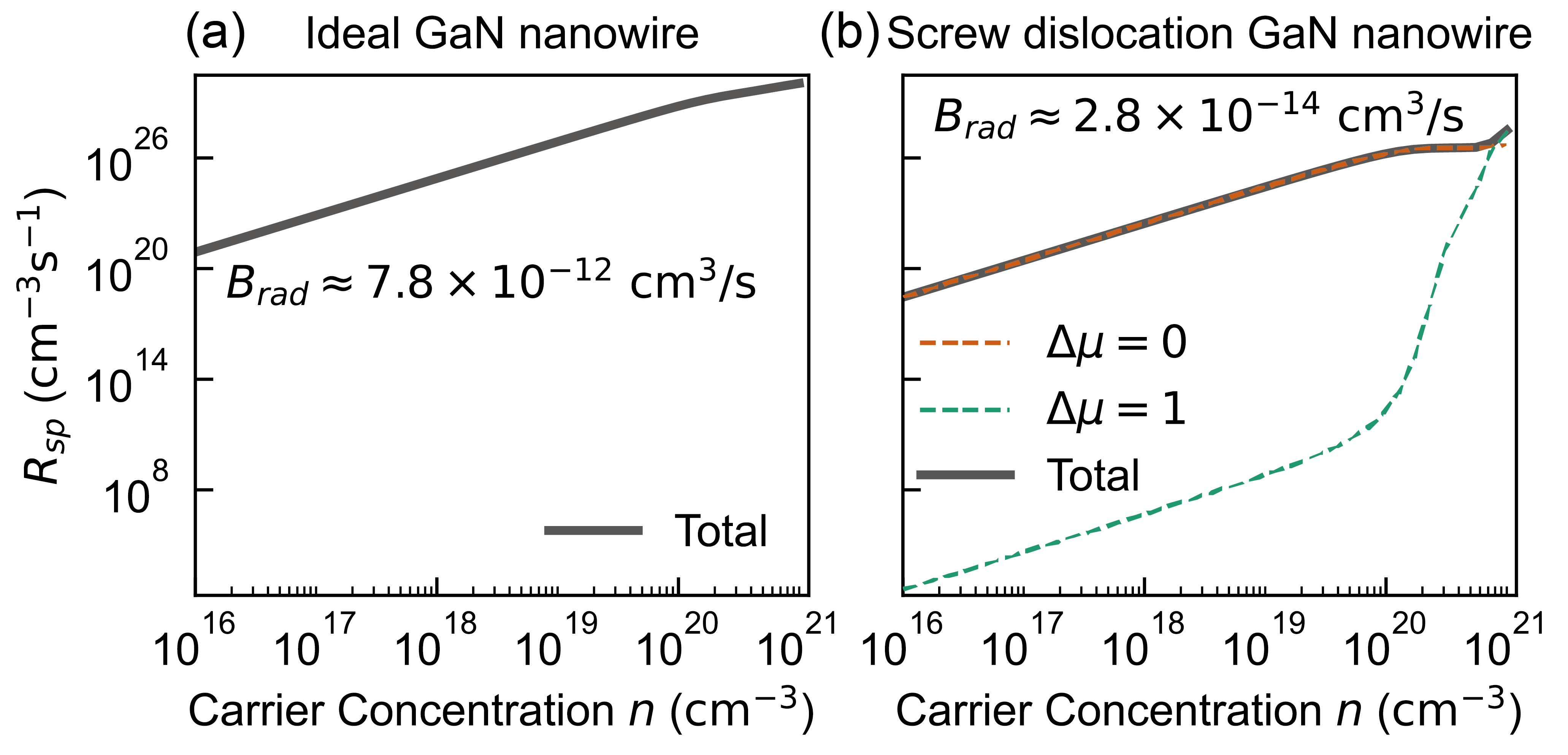}
\caption{Radiative recombination rate as a function of injected carrier density $n$. The decomposition into different $\Delta \mu$ channels is indicated by dashed lines in different colors for (a) ideal GaN and (b) GaN with a screw dislocation. \textcolor{red}{This} comparison shows the existence of screw dislocation significantly suppresses the radiative ability of GaN.}
\label{fig:rad_rates}
\end{figure}

Fig.~\ref{fig:rad_rates} presents the computed radiative rates. We observe that in the GaN nanowire with a screw dislocation, $R_{\mathrm{rad}}$ is dominated by the $\Delta \mu=0$ channel, corresponding to states 3 and 4 at the $\Gamma$ point at low carrier densities $n$. This corresponds to a photon energy of approximately 0.7 eV in the infrared region, which is predicted to exhibit polarization parallel to the screw dislocation line.

Compared to the ideal GaN nanowire, the GaN nanowire containing screw dislocations exhibits a reduction in the radiative recombination rate by two to three orders of magnitude. To elucidate the microscopic origin of this phenomenon, we examine the electrostatic potential distribution and the atomic composition of the states that dominate radiative recombination.

\begin{figure}[ht]
\centering
\includegraphics[width=1.05\columnwidth]{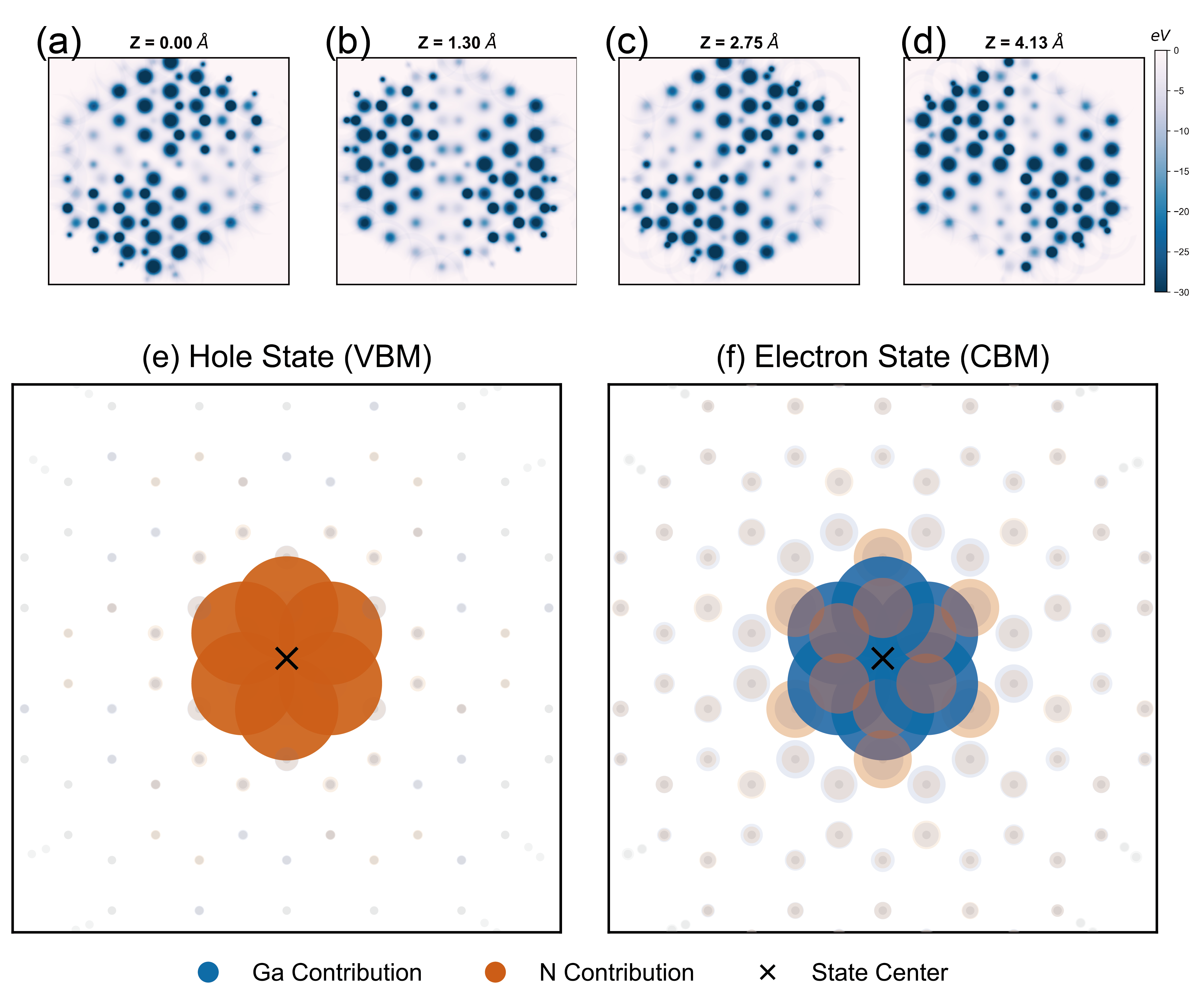}
\caption{Electrostatic potential distribution and orbital composition analysis of screw-dislocated GaN at the $\Gamma$ point.
(a)-(d) Cross-sectional maps of the electrostatic potential at different $z$-coordinates ($z=0.00, 1.30, 2.75,$ and $4.13$ \AA). 
(e)-(f) Orbital compositions of states 3 and 4: The marker size represents the orbital weight evaluated via Mulliken population analysis, and colors represent atom types. The hole state is predominantly localized on nitrogen atoms, while the electron state is localized on gallium atoms.}
\label{fig:pot}
\end{figure}

\subsection{Physical Interpretation}
\label{physical_interpretation}
To elucidate the microscopic origin of the precipitous suppression of the radiative recombination rate, we identify a synergistic mechanism combining GaN's strong piezoelectric response with a local quantum confined effect induced by the screw dislocation~\cite{takeuchi1997quantum, ryou2009control, zhu2017influence}.

Due to the large piezoelectric coefficients of GaN, the severe strain field of the screw dislocation generates a substantial internal electrostatic potential. As illustrated in Fig.~\ref{fig:pot}(a)-(d), this potential exhibits a strong angular inhomogeneity within the $xy$-plane, characterized by distinct local minima and maxima. Crucially, due to the screw symmetry, this non-uniform potential distribution rotates along the $z$-axis. This spiraling potential landscape creates strong local electric fields along the $z$-direction, acting as a driving force for charge separation.

This internal field effectively transforms the dislocation core into a "quasi-quantum well" where charge carriers are spatially pulled apart. This spatial segregation is explicitly confirmed by the orbital composition analysis of the gap states near the Fermi level, as shown in Fig.~\ref{fig:pot}(e) and (f). While both State 3 (below $E_F$) and State 4 (above $E_F$) are localized around the dislocation core, they exhibit distinct atomic origins: State 3 is predominantly distributed on Nitrogen (N) atoms, whereas State 4 is localized mainly on Gallium (Ga) atoms.

From a quantum-mechanical perspective, this spatial mismatch between the electron (Ga-dominated) and hole (N-dominated) wavefunctions drastically reduces the overlap integral. Consequently, the transition dipole matrix element $\mathbf{d}_{fi}= -e\int\psi_f^*(\mathbf{r})\,\mathbf{r}\,\psi_i(\mathbf{r})\,d^3r$ vanishes because the integrand becomes negligible where the two wavefunctions do not co-locate. Since the spontaneous-emission rate scales as $|\mathbf{d}_{fi}|^2$, this separation leads directly to the observed reduction in radiative recombination by orders of magnitude.

\subsection{Implications for dislocation-related luminescence}
Applying the selection rule \eqref{eq:selection_rule} to the dislocation-localized states identified in the electronic-structure calculation immediately constrains both capture and emission pathways. For example, an optically allowed radiative recombination from a conduction-band state in block $\mu_c$ to a dislocation state in block $\mu_D$ requires $\mu_D-\mu_c\equiv m\pmod 6$, and the emitted photon polarization then reflects the corresponding $m$ channel.

In the specific case of GaN, our analysis indicates that the dominant radiative transitions originate from the dislocation-localized states 3 and 4. These transitions occur in the infrared spectral range and correspond to the $\Delta \mu = 0$ channel. This implies that the emitted photons are linearly polarized parallel to the screw dislocation line. Consequently, the polarization anisotropy of this infrared luminescence offers a promising spectroscopic method to probe the spatial orientation and distribution of screw dislocations in bulk GaN.

\section{Nonradiative recombination}
This section distills the essential results for phonon-mediated
nonradiative recombination in the screw-dislocated GaN. Given that the energy gap between band 3 and 4 is the smallest, the multi-phonon non-radiative process is expected to be most efficient and thus primarily takes place between these two dislocation states.

The nonadiabatic coupling driving transitions between electronic channels $i\to f$ is given by the off-diagonal pieces of phonon Hamiltonian $\hat{H_{el}}$. Projected onto single-channel nuclear eigenstates $\Psi_{i,\nu}(Q)$ (solutions of the diagonal
channel nuclear equation), the phonon-mediated transition rate between channel eigenstates follows from Fermi's golden rule~\cite{huang1981lattice}:
\begin{equation}\label{eq:FGR_nonrad}
W_{i\nu\to f\nu'} = \frac{2\pi}{\hbar} \lvert M_{fi;\nu'\nu} \rvert^2 \; \delta(\mathcal{E}_{f,\nu'}-\mathcal{E}_{i,\nu}),
\end{equation}
where the matrix element is~\cite{passler1974description,passler1975description}
\begin{equation}
\label{eq:Mfi_general}
    M_{fi;\nu'\nu}=\int dQ\;\Psi^*_{f,\nu'}(Q)\hat{H_{el}}\Psi_{i,\nu}(Q).
\end{equation}

For thermally populated initial nuclear states one obtains the total (channel-to-channel) rate
\begin{equation}\label{eq:W_total}
W_{i\to f}(T) = \sum_{\nu,\nu'} P_{i,\nu}(T)\,W_{i\nu\to f\nu'}
\end{equation}
with Boltzmann weights $P_{i,\nu}(T)=e^{-\beta\mathcal{E}_{i,\nu}}/Z_i(T)$. The macroscopic nonradiative recombination
rate follows by summing over defect concentrations and initial carrier populations as discussed earlier.

\begin{figure}[ht]
\centering
\includegraphics[width=1.0\columnwidth]{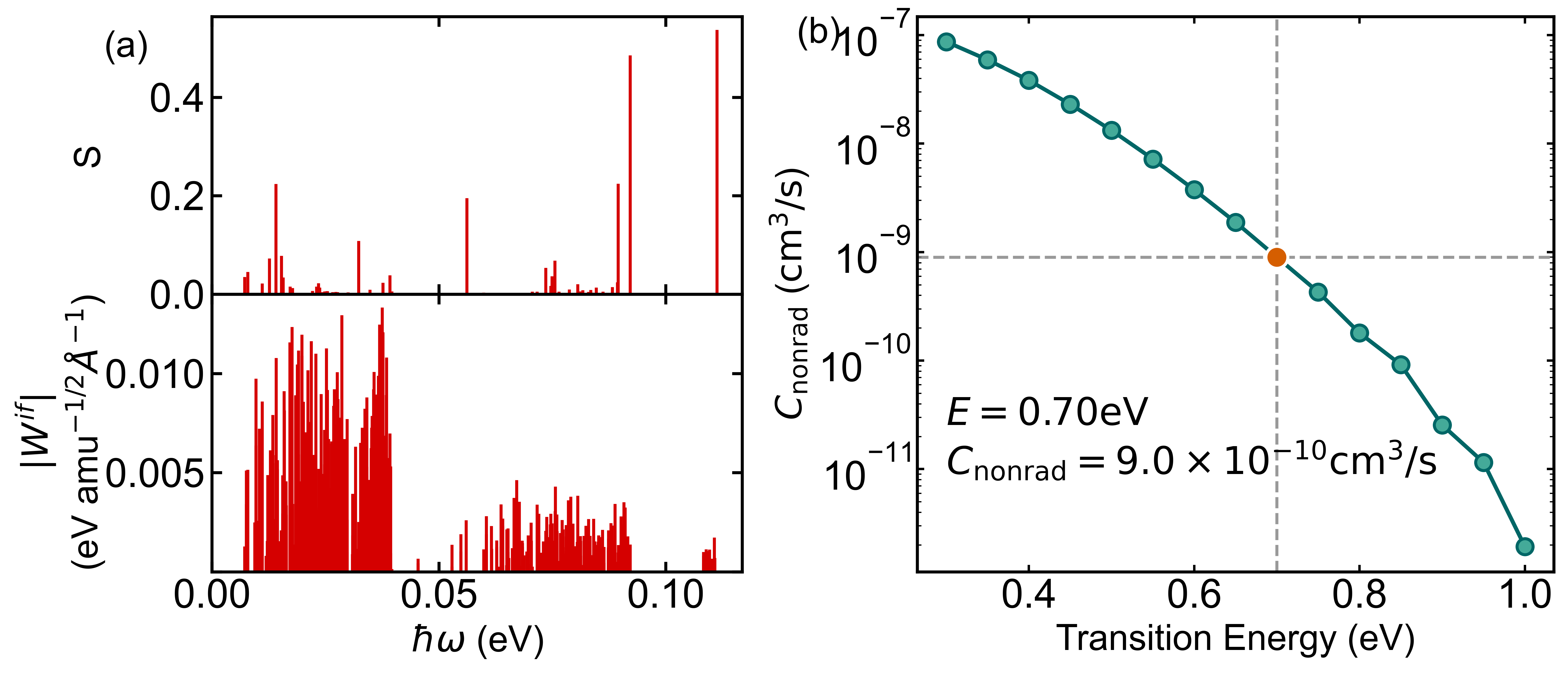}
\caption{The contribution of different phonon modes to the Huang-Rhys factor ($S$), electron-phonon coupling matrix
element ($W^{if}$), and nonradiative capture coefficient ($C_{\text{nonrad}}$). $S$ involves contributions from phonons across the entire frequency spectrum and the low-frequency regime dominates the electron-phonon coupling.}
\label{fig:nonrad}
\end{figure}

Calculation shows there is a lattice relaxation of the screw dislocation, i.e., the equilibrium structures differ before and after nonradiative recombination. We calculate the partial Huang-Rhys factor ($S_k=\omega_k\Delta Q_k^2/2\hbar$)~\cite{huang1950theory} at each phonon mode $k$ to reflect the strength of lattice relaxation, as shown in the upper panel of Fig.~\ref{fig:nonrad}. The total Huang-Rhys factor of 2.6 can be obtained by summing $S_k$ at each mode, which is much lower than typical $S$ for carrier capture at point defects in GaN, e.g., $S=10.0$ for hole capture at $C_\text{N}^-$~\cite{alkauskas2014first}. Furthermore, the phonon modes with large contribution to $S_k$ contribute minorly to the electron-phonon coupling matrix element (Fig.~\ref{fig:nonrad}). Since the recombination rate is determined by a joint contribution of lattice relaxation and electron-phonon coupling ($W^{if}$)~\cite{zhou2025defect}, such a mismatch will also reduce the practical contribution to this multiphonon process. As a result, the calculated nonradiative recombination is limited, with a coefficient at energy gap $\Delta E = 0.7\,\text{eV}$ and $T= 300\,\text{K}$,
\begin{equation}
    C_{\text{nonrad}}=9.0\times 10^{-10}\,\text{cm}^3/\text{s}.
\end{equation}
Although $S$ of the screw dislocation is smaller than that of $C_\text{N}^-$, the calculated recombination coefficient is still comparable to that of $C_\text{N}^-$ in Ref.~\cite{alkauskas2014first} ($C_p = 3.1 \times 10^{-9}\,\text{cm}^3/\text{s}$), which can be ascribed to the smaller energy gap in our nonradiative recombination calculations. The calculated coefficient is consistent with the reduced recombination lifetimes observed in dislocation-rich GaN~\cite{gao2025first}.

In contrast, the radiative coefficient is $B_{\text{rad}}=2.8\times 10^{-14}\,\text{cm}^3/\text{s}$. Thus nonradiative processes dominate by many orders of magnitude under the considered conditions, implying that screw dislocations can severely quench photoluminescence and adversely affect the performance of GaN-based light-emitting devices.

\section{Conclusion}
In this paper, we have developed a compact and unified framework for analyzing electronic structure and
radiative/nonradiative processes in crystals that host screw dislocations. Starting from the screw symmetry
generator we constructed localized atomic orbital Bloch sums and discrete-group projectors that produce
symmetry-adapted subspaces labeled by the screw representation index $\mu$. When the Hamiltonian commutes with the screw operator the projector construction yields an exact block decomposition
\(\hat H=\bigoplus_{\mu}H_{\mu}(k)\) so that each block can be diagonalized independently to obtain
symmetry-resolved bands $\epsilon_{n,\mu}(k)$. This approach avoids the inefficiencies of plane-wave supercell expansions for localized defect cores while preserving the full group-theoretic content of the problem.

Applying the method to GaN with a sixfold screw generator we proved a robust band-flow constraint
($\Delta\mu=+2$ mod 6) that enforces how bands from different $\mu$-blocks connect across the axial Brillouin-zone.
The projector-based decomposition clarifies band assignments, enables polarization-resolved optical matrix-element
analysis, and directly yields the selection rule $\Delta\mu\equiv m\pmod{6}$ for electric-dipole transitions. 
These results give a transparent explanation for polarization-dependent selection channels and for how dislocation
localized states couple to extended bands.

Using calculated Hamiltonians and optical matrix elements we computed radiative rates and the dielectric response resolved by $\mu$-channel. Our numerical results indicate a
strong suppression (2–3 orders of magnitude) of radiative recombination in the presence of screw cores; physically
this is attributable to a piezoelectrically driven field that spatially separates electron and hole
densities and therefore quenches the transition dipole. Finally, by streamlining the Born–Huang projection and derivative-coupling formalism, we evaluate the nonradiative recombination coefficient and find that it exceeds the radiative counterpart by several orders of magnitude, indicating that only a small fraction of excited carriers contributes to photon emission.

The symmetry-adapted block decomposition and the derived selection rules form a foundation for several follow-up
studies: (i) generation of quantitative radiative and nonradiative lifetimes from converged DFT inputs and comparison with experiment; (ii) construction of reduced-order models for carrier transport along dislocation lines
that incorporate $\mu$-resolved scattering channels; (iii) systematic exploration of other screw symmetries and
their mathematically constrained coupling chains. We anticipate that the formal and computational tools presented here
will aid interpretation of dislocation-related luminescence and carrier-loss mechanisms in wide-bandgap semiconductors
and will generalize readily to other materials with screw structural motifs.

\acknowledgments
This work was supported by the National Natural Science Foundation of China (Grants No. 12188101, No. 12334005, No. 12404089, No. 1227408, No. 124B1003), National Key Research and Development Program of China (Grant No. 2024YFA1409800).

The data that support the findings of this study are available within the article and its Supplemental Material as cited in Ref.~\cite{data_reference}.

Y.X. and H.S. contributed equally to this work.
\bibliography{ref}

\end{document}